\documentclass[preprint,prd,aps,showpacs,showkeys,nofootinbib]{revtex4}
\usepackage{amssymb}
\usepackage{amsmath}
\usepackage{graphicx}
\usepackage{subfigure}
\usepackage{float}
\usepackage{mathrsfs}

\usepackage{amsmath}
\usepackage{epstopdf}
\usepackage{dcolumn}
\usepackage{bm}
\usepackage{subfig}
\usepackage{comment}

\newcommand{\be}{\begin{eqnarray}}
\newcommand{\ee}{\end{eqnarray}}

\begin{document}


\title{Nanohertz gravitational waves from domain walls nucleated during inflation}

\author{Zhi-Yong Huang}
\email{huangzhiyong@stu.xidian.edu.cn}
\author{Tie-Jun Gao}
\email{tjgao@xidian.edu.cn}
\affiliation{School of Physics, Xidian University, Xi'an 710071, China}

\begin{abstract}
We investigate scalar-induced gravitational waves (SIGWs) produced by domain walls (DWs) nucleated via quantum tunneling during inflation with an extended nucleation time. In contrast to the small-period nucleation framework, in which DWs form over a narrow interval and possess nearly identical radii, we show that an extended nucleation period leads to a distribution of DW radii characterized by $\gamma\equiv\overline{R^{10}}/(\overline{R^5})^2>1$, which enhances the resulting curvature perturbations. We construct a two-field inflation model with an inflaton $\phi$ and a spectator field $\chi$ coupled through the potential $V(\phi,\chi)$, where the DW tension $\sigma(t)$ evolves smoothly as the inflaton rolls past a critical value. The characteristic width of this transition determines the cutoff scale $k_{\text{cut}}$ of the curvature power spectrum, enabling the SIGW peak to be placed in the nanohertz frequency band with detectable amplitude. For three representative parameter sets, we compute the SIGW spectra and find that the peak frequency ranges from $\sim10^{-8}$\,Hz to $\sim10^{-1}$\,Hz, with the nanohertz-peaked spectrum matching the NANOGrav and EPTA signals for suitable parameter choices. By selecting different parameters, our model can produce signals detectable by different gravitational-wave detectors.

\end{abstract}

\maketitle

\section{Introduction \label{sec:intr}}
The pulsar timing array (PTA) collaborations — including NANOGrav, EPTA, PPTA, and CPTA — have observed a common-spectrum process \cite{NANOGrav:2023,EPTA:2023,PPTA:2023a,CPTA:2023,Shannon:2015,Lentati:2015,NANOGrav:2018,EPTA:2023a} consistent with the Hellings–Downs angular correlation \cite{Hellings:1983}, strongly supporting its gravitational-wave origin. This detection of a stochastic gravitational-wave background (SGWB) in the nanohertz frequency band has opened a new window into the early Universe \cite{NANOGrav:2023,BurkeSpolaor:2019,Jenet:2006,Verbiest:2016}. Candidate sources of this signal are actively investigated, including gravitational waves from supermassive black-hole binaries, from topological defects such as cosmic strings and DWs, from first-order phase transitions, and from scalar-induced mechanisms (SIGWs) \cite{NANOGrav:2023b,Ferreira:2024}.

Domain walls (DWs) are sheet-like topological defects that arise when a discrete symmetry is spontaneously broken \cite{Zeldovich:1974,Kibble:1976}. A DW separates spatial regions occupying different degenerate vacua, with the scalar field interpolating between adjacent vacua across the interface. The energy stored in a DW is characterized by its tension $\sigma$ \cite{Vilenkin:1985}, and the network dynamics are governed by the competition between tension-driven collapse and Hubble friction. If the discrete symmetry is exact, the resulting DW network would eventually dominate the energy density of the Universe, leading to the well-known domain-wall problem; this is avoided if the symmetry is explicitly broken by a small bias term, causing adjacent DWs to annihilate and release their stored energy as gravitational waves \cite{Zeldovich:1974,Saikawa:2017,Hiramatsu:2014,Vilenkin:1981,Vachaspati:1984,Pujolas:2023,Blasi:2023}. Beyond this conventional annihilation channel, DWs can also generate SIGWs: their inhomogeneous spatial distribution sources density perturbations that induce secondary tensor modes upon horizon re-entry.

A particularly compelling scenario involves DWs nucleated during inflation via quantum tunneling \cite{BGV1991}. In the conventional thermal mechanism, DWs form when the Universe cools through a critical temperature and degenerate vacua are randomly selected across causal patches (Kibble mechanism) \cite{Kibble:1976}, resulting in a percolating network that pervades the entire Hubble volume. By contrast, during inflation, Hubble friction can keep the field near the top of the potential, and DWs nucleate as spherical bubble walls when the field quantum-tunnels between the two degenerate vacua \cite{BGV1991}. This inflationary nucleation scenario avoids the domain-wall problem: the nucleation rate $\lambda \propto e^{-S_E}$ is strongly suppressed by the Euclidean action $S_E$, limiting the total number of DWs; each nucleated DW is stretched beyond the Hubble horizon by inflation, reducing its energy density; and after inflation ends, DWs re-enter the horizon and collapse into radiation rather than forming a scaling network \cite{Vachaspati:1984,Deng:2017}, so DW domination is avoided without needing a bias term \cite{BGV1991,Garriga:1994}. In this work, we propose an extended-period nucleation model in which the DW tension $\sigma(t)$ varies smoothly during inflation, allowing DWs to nucleate over an extended period. This is achieved by introducing a width parameter $\Delta$ in the scalar potential, which controls the profile of $\sigma(t)$. As a result, DWs acquire a distribution of comoving radii, quantified by the factor $\gamma \equiv \overline{R^{10}}/(\overline{R^5})^2 > 1$, which enhances the variance of the density perturbation. Furthermore, the cutoff scale $k_{\mathrm{cut}}$ of the curvature power spectrum becomes tunable through the nucleation history, enabling us to place the induced gravitational-wave peak in the nanohertz band.

We work within an inflationary model where the inflaton $\phi$ and a spectator field $\chi$ interact through the potential $V(\phi, \chi)$, which contains two degenerate vacua in the $\chi$-direction. Solving the background dynamics and the nucleation rate, we compute the SIGW energy density and find that the resulting spectrum can peak in the nanohertz band, providing a viable explanation for the observed PTA signal, while the model simultaneously predicts testable signatures at higher frequencies accessible to future space-based interferometers.

The remainder of this paper is organized as follows. In Sec.~\ref{sec:DW}, we review the mechanism of gravitational wave production from domain walls formed via quantum tunneling, including the nucleation statistics, the resulting curvature perturbations, and the formalism of scalar-induced gravitational waves. In Sec.~\ref{sec:model}, we present our model with an extended nucleation time, derive the modified DW statistics, and obtain the numerical evolution of the wall network. In Sec.~\ref{sec:SIGW}, we compute the scalar-induced gravitational wave spectrum and compare our predictions with PTA observations. We summarize our findings in Sec.~\ref{sec:summary}.

\section{SPHERICAL DW NUCLEATION AND EVOLUTION \label{sec:DW}}
A DW is a two-dimensional topological defect that arises when a discrete symmetry is spontaneously broken \cite{Zeldovich:1974,Kibble:1976}. Consider a real scalar field $\chi$ with the $\mathbb{Z}_2$-symmetric potential
\begin{equation}
V(\chi)=\frac{\lambda}{4}(\chi^2-\upsilon^2)^2\,,
\end{equation}
which possesses two degenerate minima at $\chi=\pm\upsilon$. The static planar DW solution interpolating between the two vacua is
\begin{equation}
	\chi(x)=\upsilon\tanh\left(\sqrt{\frac{\lambda}{2}}\,\upsilon\,x\right),
\end{equation}
and the associated surface energy density (tension) is
\begin{equation}
	\sigma=\frac{4}{3}\sqrt{\frac{\lambda}{2}}\,\upsilon^3.
\end{equation}

In a cosmological context, if the discrete symmetry is exact, the DW network energy density scales as $\rho_{\text{DW}}\propto a^{-1}$, eventually dominating over radiation and matter — the well-known domain-wall problem. This is resolved by introducing a small bias term that explicitly breaks the symmetry, causing adjacent walls to annihilate and release gravitational waves \cite{Zeldovich:1974,Saikawa:2017}.

\begin{figure}[htbp]
	\centering
	\includegraphics[width=0.8\textwidth]{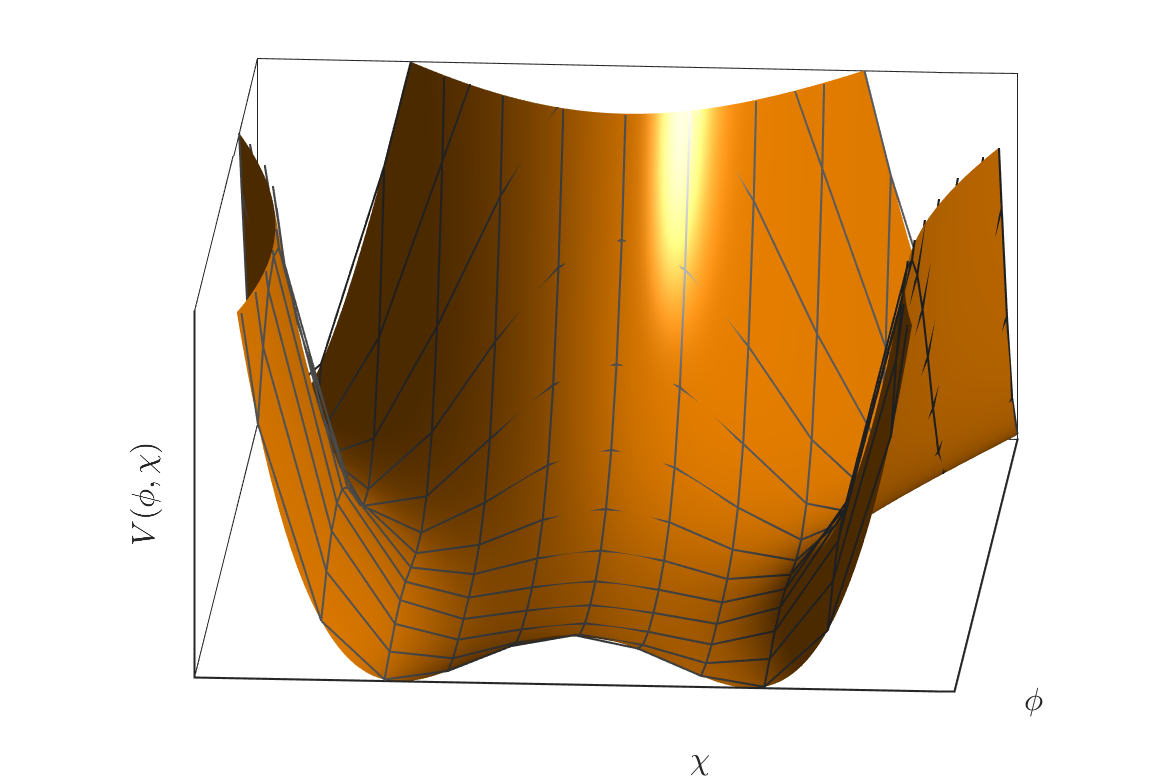}
	\caption{Schematic of the potential $V(\phi,\chi)$, where the $\chi$-direction barrier height evolves with $\phi$.}
	\label{fig:Vphichi}
\end{figure}

Beyond the Kibble mechanism, DWs can also nucleate during inflation via quantum tunneling. To illustrate this mechanism, we consider an example involving the inflaton $\phi$ and a spectator field $\chi$, coupled through the potential
\begin{equation}
	V(\phi,\chi)=\frac{\lambda_\chi}{4}\bigl[\chi^2-\alpha^2(\phi-\phi_c)^2-m^2\bigr]^2+f(\phi), 
	\label{eq:V}
\end{equation}
as introduced in Refs.~\cite{Zeng:2023,Liu:2019}. 
The $\chi$-direction supports two degenerate vacua separated by a barrier whose height depends on $\phi$, as shown in Fig.~\ref{fig:Vphichi}. At the onset of inflation, the inflaton $\phi$ starts from one side of its potential valley and rolls slowly along its potential during inflation; as it approaches $\phi_c$, the barrier height in the $\chi$-direction decreases. Once the barrier becomes sufficiently low, quantum tunneling between the two degenerate vacua becomes efficient, and bubbles of one vacuum nucleate inside the other; these nucleation events are rare and independent, and each bubble is bounded by a closed DW.  

The nucleation rate per unit physical volume per unit time is given by
\begin{equation}
	\lambda(t)=H^4(t)Ae^{-S_E(t)}, 
	\label{eq:nucleation}
\end{equation}
which is derived within the semiclassical approximation. The semiclassical expression is used in the regime $S_E\gtrsim1$, where the exponential factor in Eq.~\eqref{eq:nucleation} suppresses the local nucleation rate; the semiclassical expansion is under parametric control for $S_E\gg1$. For $S_E\ll1$, the dilute-nucleation description ceases to apply, and prolific production may lead to a percolating DW network. Once $S_E$ is larger than order unity, the local rate is exponentially suppressed; whether the resulting population is sparse, however, depends on the full nucleation history rather than on $S_E$ alone. For our benchmark parameters, the Euclidean action satisfies $S_E\gtrsim20$ throughout the nucleation period. The nucleation results summarized in Table~\ref{tab:2} indicate that the nucleated DWs are sparse and can be treated as isolated spherical walls. Within this model, nucleation events in different Hubble patches are treated as independent, consistent with the Poisson statistics adopted in Sec.~\ref{sec:model}. In the rate formula, $A$ represents a dimensionless prefactor which varies slowly with the ratio $\sigma H^{-3}$; its magnitude is typically of order unity, $A \sim 1$ \cite{BGV1991,Garriga:1994}. The Euclidean action $S_E$ can be expressed as
\begin{equation}
	S_E(t)=2\pi^2\sigma(t)H^{-3}(t). 
	\label{eq:SE}
\end{equation}
The parametric form $S_E\sim\sigma/H^3$ follows from dimensional analysis, and the coefficient $2\pi^2$ is the exact thin-wall value \cite{BGV1991}. The benchmark parameters adopted for the numerical analysis below lie well inside the thin-wall regime: in terms of the thin-wall parameter $x\equiv\sqrt{2\lambda_\chi}\,\upsilon/H$, where $\upsilon$ is the vacuum expectation value of $\chi$, the dimensionless wall thickness satisfies $\delta_wH\equiv 2/x\lesssim0.2353$ throughout the nucleation period (see Table~\ref{tab:2}). 

DWs nucleated during inflation are stretched superhorizon and later re-enter as subhorizon perturbations. Because nucleation events in different Hubble patches are independent, the DW number density follows Poisson statistics, which generates isocurvature density perturbations. In the long-wavelength limit, these isocurvature fluctuations are converted into an adiabatic curvature perturbation that remains conserved on superhorizon scales; the density contrast is then related to the conserved curvature perturbation by the standard radiation-era matching relation at horizon crossing, $P_\delta(k)=\frac{16}{81}\mathcal{P}_{\mathcal{R}}(k)$, yielding a characteristic power spectrum $\mathcal{P}_{\mathcal{R}}(k)\propto k^3$ \cite{BBKS:1986,Zeng:2023}. These curvature perturbations source SIGWs at second order. The induced gravitational waves are computed from the second-order tensor perturbation. In the Newtonian gauge, the equation of motion for the transverse-traceless tensor mode $h_{\boldsymbol{k}}$ is
\begin{equation}
	h_{\boldsymbol{k}}''+2\mathcal{H}h_{\boldsymbol{k}}'+k^2h_{\boldsymbol{k}}=4S_{\boldsymbol{k}}(\tau), 
	\label{eq:7}
\end{equation}
where $S_{\boldsymbol{k}}$ is a quadratic source term constructed from the scalar perturbations. The solution is obtained via the Green's function method,
\begin{equation}\label{eq:hk_Green}
	h_{\boldsymbol{k}}(\tau)=\frac{4}{a(\tau)}\int^{\tau}d\bar{\tau}a(\bar{\tau})G_k(\tau,\bar{\tau})S_{\boldsymbol{k}}(\bar{\tau}). 
\end{equation}
The present-day gravitational-wave energy density spectrum $\Omega_{\text{GW}}(f)$ is then obtained by integrating over the curvature power spectrum with appropriate kernel functions and thermal-history correction factors. The detailed calculation is presented in Sec.~\ref{sec:SIGW}.

The above framework was implemented in Ref.~\cite{Zeng:2023} using a Starobinsky-type inflaton potential $f(\phi)=\Lambda_0(1-e^{-\sqrt{2/3}\,\phi})^2$~\cite{Starobinsky:1980}. In their model, the Euclidean action $S_E$ exhibits a sharp spike near $\phi=\phi_c$, forcing all DWs to nucleate within a very narrow time interval. Consequently, all DWs possess nearly identical comoving radii, and the amplitude of $\mathcal{P}_\mathcal{R}$ is essentially fixed by the peak value of $S_E$. The resulting curvature power spectrum is therefore sharply peaked, and the predicted GW spectrum lies in the millihertz band. In the extended-period nucleation model proposed in the present work, a smooth $\sigma(t)$ profile extends the nucleation period and gives the DWs a broad distribution of radii. This enhances the curvature power spectrum and shifts the SIGW peak into the nanohertz band, where, for suitable parameter choices, the signal can match the PTA observations; different parameter choices instead retain signals at millihertz to decihertz frequencies.

\section{DWs nucleation model with extended-period nucleation \label{sec:model}}
\subsection{Extended-period nucleation statistics of DWs}
In this section, we will construct a DWs nucleation model based on the extended-period nucleation under the quantum tunneling effect. The nucleation of quantum topological defects during inflation has been studied in past literatures. The Euclidean version of de Sitter space is geometrically a four-sphere whose radius is given by the inverse Hubble parameter, $H^{-1}$. Through quantum tunneling, a DW can nucleate during inflation. In the Euclidean picture, the nucleation process is described by an instanton solution, which takes the form of a three-sphere of the same maximal radius, $H^{-1}$, embedded within this four-sphere. The nucleation rate and Euclidean action are given by Eqs.~\eqref{eq:nucleation} and \eqref{eq:SE}, respectively. 

As discussed in Sec.~\ref{sec:DW}, the small-period nucleation framework assumes all DWs possess nearly identical radii. In the extended-period nucleation scenario, DWs nucleate over an extended period, acquiring a distribution of comoving radii $R$. The number density of DWs nucleated between $t_1$ and $t_2$ is
\begin{equation}
	n = \int_{t_1}^{t_2} \lambda(t) \, a^3(t) \, dt, 
	\label{eq:9}
\end{equation}
the integration interval $[t_1, t_2]$ covers the main period of DW nucleation. 


\subsection{Superhorizon evolution and horizon re-entry of DWs}
Since DWs nucleate over an extended period, each DW acquires, at its nucleation time $t_i$, a comoving radius $R(t_i) = [a(t_i) H(t_i)]^{-1}$; as a result, different DWs re-enter the horizon at different times. The $i$-th DW re-enters the horizon when its physical radius equals the Hubble radius, $a(t_{r,i})\, H(t_{r,i})\, R(t_i) = 1$, where $t_{r,i}$ denotes its horizon re-entry time. We denote by $t_e$ the end of inflation, and assume a short reheating so that the Universe enters the radiation-dominated (RD) era soon after inflation. During the RD era, the scale factor scales as $a(t) \propto t^{\frac{1}{2}}$, and the Hubble parameter scales as $H(t) \propto (2t)^{-1}$; the combination $a^2 H = a^2(t_e) H(t_e)$ is constant. The horizon re-entry time and the corresponding scale factor of the $i$-th DW are then given by
\begin{equation}
	t_{r,i}^{\frac{1}{2}} = t_e^{\frac{1}{2}} \, a(t_e) \, H(t_e) \, R(t_i),
	\label{eq:tr}
\end{equation}
\begin{equation}
	a(t_{r,i}) = a^2(t_e) \, H(t_e) \, R(t_i).
	\label{eq:atr}
\end{equation}

Averaging over the nucleation distribution, the energy density of DWs at the end of inflation is given by
\begin{equation}
	\overline{\rho}_{\text{DW}} = \frac{4\pi \sigma(t_e)}{a(t_e)} \int_{t_1}^{t_2} \lambda(t_i) \, a^3(t_i) \, R^2(t_i) \, dt_i.
\end{equation}

It can be seen that if the tension of DWs remains constant after inflation, then on superhorizon scales, the energy density of spherical DWs scales as $\overline{\rho}_{\text{DW}} \propto a^{-1}$, while the total energy density in the RD era scales as $\rho_{\text{tot}} \approx \rho_r \propto a^{-4}$ (with $\rho_{\text{tot}} = 3H^2(t_e)$). Consequently, the ratio $\overline{\rho}_{\text{DW}}/\rho_r$ grows as $a^3$. Weighting each DW by its conversion factor $\left( a(t_{r,i})/a(t_e) \right)^3$ and summing over the nucleation distribution, we obtain the converted energy fraction
\begin{equation}
	\left. \frac{\overline{\rho}_{\text{DW}}}{\rho_r} \right|_{tr} = \frac{4\pi\sigma(t_e)}{3a(t_e)H^2(t_e)} \int_{t_1}^{t_2} \lambda(t_i) \, a^3(t_i) \, R^2(t_i) \left( \frac{a(t_{r,i})}{a(t_e)} \right)^3 dt_i.
\end{equation}

Note that even inside a Hubble patch containing a spherical DW, $\overline{\rho}_{\text{DW}}$ cannot exceed $\rho_r$. Otherwise, the patch would collapse into a PBH before the DW re-enters the horizon. This condition imposes $\overline{\rho}_{\text{DW}}/\rho_r < p$ \cite{Deng:2017}. The probability $p$ of nucleating one spherical domain wall within a Hubble volume is given by
\begin{equation}
	p = \int_{t_1}^{t_2} \frac{4}{3}\pi \left( \frac{1}{H(t)} \right)^3 \lambda(t) \, dt,
	\label{eq:p}
\end{equation}
If the interaction between DWs and matter fields is non-negligible, spherical DWs dissipate their energy into the background radiation upon horizon re-entry, thereby generating density perturbations.


\subsection{Curvature perturbations from Poisson-distributed DWs}
The nucleation of DWs is treated as an independent and random process, which naturally follows a Poisson distribution. Each DW's comoving radius $R(t_i)$ is fixed at its nucleation time. 

To ensure a meaningful statistical sample, we consider a comoving volume of $(2L)^3$, where L is chosen to be larger than the characteristic comoving separation of the DWs, denoted as $S_{\text{cut}}$. As $L$ approaches $S_{\text{cut}}$ from above, the mean number of DWs within the smoothing volume drops to order unity. The separation $S_{\text{cut}}$ is the radius of a sphere within which, on average, exactly one DW is contained, i.e., $\frac{4}{3}\pi S_{\text{cut}}^3 n = 1$.

The variance of the density perturbation after smoothing on a scale $L$ \cite{BBKS:1986,Zeng:2023} is given by
\begin{equation}
	\sigma_{\delta}^{2}(L) = \int d\ln k \, P_{\delta}(k) \exp\left( -k^{2} L^{2} \right), 
	\label{eq:10}
\end{equation}

where the power spectrum of density perturbations is defined as $P_\delta(k) \equiv \frac{k^3}{2\pi^2} |\delta_{\boldsymbol{k}}|^2$. Here $\delta_{\boldsymbol{k}}$ is the Fourier transform of the density contrast $\delta(\boldsymbol{x})\equiv\delta\rho(\boldsymbol{x})/\rho$. 

The total density perturbation is given by
\begin{equation}
	\delta_{\text{tot}} = \frac{\delta\rho_r + \delta\rho_{\text{DW}}}{\rho_r + \rho_{\text{DW}}},
	\label{eq:11}
\end{equation}
Since current observations show that DWs did not dominate the Universe, $\rho_{\rm DW} \ll \rho_r$, the denominator of Eq.~\eqref{eq:11} can be approximated by $\rho_r$. Meanwhile, $\delta\rho_r$ originates from vacuum fluctuations during inflation, so that $\delta\rho_r/\rho_r \sim 10^{-5}$, while $\delta\rho_{\rm DW}$ arises from the random distribution of spherical DWs; as verified below, $\delta\rho_{\rm DW}/\rho_r$ on the scales of interest exceeds $10^{-5}$, i.e., $\delta\rho_{\rm DW} > \delta\rho_r$ \cite{Zeng:2023}. In this case the density perturbation is dominated by the DW component, and Eq.~\eqref{eq:11} reduces to
\begin{equation}
	\delta_{\rm tot} \approx \frac{\delta\rho_{\rm DW}}{\rho_r}.
	\label{eq:12}
\end{equation}

Assuming that the number of DWs contained within a comoving volume of $(2L)^3$ is $X$, and each DW has a radius $R(t_i)$ at its nucleation time $t_i$, where $t_e$ is the end of inflation, the total energy of DWs within the volume is given by
\begin{equation}
	E_{\text{DW}} = \sum_{i=1}^{X} 4\pi\sigma(t_e) a^2(t_e) R^2(t_i), 
	\label{eq:EDW}
\end{equation}
let $\overline{X}$ denote the expectation value of $X$, and $\overline{R^2}$ denote the expectation value of $R(t_i)^2$. The average energy density within the volume is then given by
\begin{align}
	\overline{\rho}_{\text{DW}} &= \frac{\overline{E}_{\text{DW}}}{(2 a(t_e) L)^3} \notag \\
	&= \frac{\overline{X} 4\pi (a^2(t_e) \overline{R^2}) \sigma(t_e)}{(2 a(t_e) L)^3},
	\label{eq:rhobarDW}
\end{align}
then the local density perturbation of DWs can be written as
\begin{align}
	\delta\rho_{\text{DW}} &= \frac{4\pi \sigma(t_e) a^2(t_e)}{\left( 2 a(t_e) L \right)^3} \Bigg[ \sum_{i=1}^{X} R^2(t_i) \left( \frac{a(t_{r,i})}{a(t_e)} \right)^3 - \overline{X}\,\overline{R^2\left( \frac{a(t_{r,i})}{a(t_e)} \right)^3} \Bigg],
	\label{eq:drhoDW}
\end{align}
the factor $\left(a(t_{r,i})/a(t_e)\right)^3$ in Eq.~\eqref{eq:drhoDW} is the conversion factor of the $i$-th DW: since the energy fraction of a superhorizon DW grows as $a^3$ from the end of inflation, each DW is weighted by this growth evaluated at its own re-entry time $t_{r,i}$. Using Eq.~\eqref{eq:atr}, the factor can equivalently be written as $[a(t_e)H(t_e)]^3R^3(t_i)$. From this, we obtain that the total density perturbation is a function of the sum of DW radii within the region
\begin{align}
	\delta_{\text{tot}} &= \frac{\pi\sigma(t_e)}{2L^3a(t_e)\rho_r} \Bigg[ \sum_{i=1}^{X} R^2(t_i) \left( \frac{a(t_{r,i})}{a(t_e)} \right)^3 - \overline{X}\,\overline{R^2\left( \frac{a(t_{r,i})}{a(t_e)} \right)^3} \Bigg] \notag \\
	&= \frac{\pi\sigma(t_e)a^2(t_e)H(t_e)}{6L^3} \left( \sum_{i=1}^{X} R^5(t_i) - \overline{X}\,\overline{R^5} \right), 
	\label{eq:16}
\end{align}


We now verify the dominance condition $\frac{\delta\rho_{\rm DW}}{\rho_r} > 10^{-5}$ assumed above. By simplification, we can obtain $\frac{\delta\rho_{\rm DW}}{\rho_r} \sim \frac{\sqrt{2}\,\pi}{3}\, L^{-3/2}\, \sigma(t_e)\, a^2(t_e)\, H(t_e)\, \sqrt{n}\, \sqrt{\overline{R^{10}}}.$ With $a(t_e) \sim 10^{25}$, $H(t_e) \sim 10^{-6}$, $\sigma(t_e) \sim 10^{-15}$, and $n \sim 10^{22}-10^{42}$. Therefore, it can be concluded that the requirements can be satisfied at the scales of interest (the input values are listed in Table~\ref{tab:2}).

For this compound Poisson process, where $X$ follows a Poisson distribution and $R(t_i)$ are independent and identically distributed random variables, the variance of the random sum $\text{Var}(\sum\limits_{i=1}^{X} R^5(t_i)) = \overline{X} \, \overline{R^{10}}.$ From this, the variance of the total density perturbation reads
\begin{align}
	\sigma_{\delta_{\text{tot}}}^2 &= \left< \delta_{\text{tot}}^2 \right> = \left( \frac{\pi\sigma(t_e)a^2(t_e)H(t_e)}{6L^3} \right)^2 \text{Var}\left( \sum_{i=1}^{X} R^5(t_i) \right) \notag \\
	&= \left( \frac{\overline{\rho}_{\text{DW}}}{\rho_r} \right)^2 \frac{1}{\overline{X}} \frac{\overline{R^{10}}}{\left( \overline{R^5} \right)^2}. 
\end{align}

Define $\gamma = \frac{\overline{R^{10}}}{\left( \overline{R^5} \right)^2}$. If all DWs have the same radius, then $\gamma = 1$; otherwise, if the radii have a distribution, $\gamma > 1$. The number of DWs $\overline{X}$ can be expressed in terms of the number density $n$ and the comoving volume as $\overline{X} = n (2L)^3$, 
\begin{equation}
	\sigma_{\delta_{\text{tot}}}^2 = \left( \frac{\overline{\rho}_{\text{DW}}}{\rho_r} \right)^2 \frac{\gamma}{n(2L)^3},
	\label{eq:18}
\end{equation}
from the above expression, it follows that the variance of the density perturbation scales as $\sigma_{\delta_{\text{tot}}}^2 \propto L^{-3}$. Furthermore, according to the relation in Eq.~\eqref{eq:10}, the density perturbation power spectrum follows a $k^3$ power law, $P_\delta(k) \propto k^3$.

At the nucleation epoch, the energy density of the Universe is dominated by the smooth inflaton, while the DWs constitute a subdominant component whose density contrast $\delta_{\rm DW} \equiv \delta\rho_{\rm DW}/\rho_{\rm DW}$ is of order unity owing to the Poisson statistics. The DW fluctuations are therefore initially of the isocurvature type, and the curvature perturbation sourced by the DWs is negligible at this stage.

This isocurvature mode is subsequently converted into an adiabatic curvature perturbation. After inflation, the Universe enters the radiation-dominated era, and the DWs stretched by inflation remain stable at superhorizon scales until they reenter the Hubble horizon at their respective reentry times $t_{r,i}$. Upon reentry, a DW begins to collapse; since the interaction between DWs and matter fields is nonnegligible, the energy stored in each DW dissipates into the background radiation at its annihilation~\cite{Vachaspati:1984,Pujolas:2023,Blasi:2023}, imprinting the Poisson-distributed spatial configuration of the DWs on the radiation fluid. After this conversion the Universe is dominated by a single component, the perturbation becomes adiabatic, and the curvature perturbation is conserved on superhorizon scales. For the modes of interest ($k \leq k_{\rm cut}$), this conservation holds until they re-enter the horizon. At horizon crossing during the radiation-dominated era, the density contrast is related to the conserved curvature perturbation by the standard matching relation~\cite{Kodama:1984,Mukhanov:1992}
\begin{equation}
	P_{\delta}(k) = \frac{16}{81} \mathcal{P}_{\mathcal{R}}(k), 
	\label{eq:19}
\end{equation}
from this formula, we can assume that $\mathcal{P}_{\mathcal{R}}(k) = A_d (k/k_{\text{cut}})^3$, where $k_{\text{cut}}$ is the cutoff scale associated with the mean DW separation. We can approximate $k_{\text{cut}}$ as $k_{\text{cut}} = S_{\text{cut}}^{-1} = (\frac{4 \pi n}{3})^{\frac{1}{3}} $. From this, Eq.~\eqref{eq:10} can be written in the following form
\begin{align}
	\sigma_\delta^2(L) &= \frac{16A_d}{81(k_{\text{cut}}L)^3} \int (kL)^2 d(kL) \exp(-k^2L^2) \notag \\
	&= \frac{4}{81} \frac{\sqrt{\pi}A_d}{(k_{\text{cut}}L)^3}, 
\end{align}
combining this with Eq.~\eqref{eq:18}, we obtain the coefficient of $\mathcal{P}_{\mathcal{R}}(k)$
\begin{align}
	A_d &= \frac{81}{32\sqrt{\pi}} \left( \frac{\overline{\rho}_{\text{DW}}}{\rho_r} \right)^2 \frac{\gamma}{n} k_{\text{cut}}^3 \notag \\
	&= \frac{27\sqrt{\pi} \gamma}{8} \left( \frac{\overline{\rho}_{\text{DW}}}{\rho_r} \right)^2. 
\end{align}

When $k > k_{\text{cut}}$, the smoothing scale $L \sim 1/k$ becomes smaller than the mean DW separation, so that the continuum description of the DW distribution breaks down. In this case, we set $\mathcal{P}_{\mathcal{R}}(k) = 0$. Therefore, the final induced curvature perturbation power spectrum is
\begin{equation}
	\mathcal{P}_{\mathcal{R}}(k) = 
	\begin{cases} 
		\dfrac{27\sqrt{\pi}}{8} \gamma \left( \dfrac{\overline{\rho}_{\text{DW}}}{\rho_r} \right)^2 \left( \dfrac{k}{k_{\text{cut}}} \right)^3 & \text{for } k \leq k_{\text{cut}}, \\
		0 & \text{for } k > k_{\text{cut}}.
	\end{cases}
	\label{eq:21}
\end{equation}


\section{SCALAR-INDUCED GRAVITATIONAL WAVES \label{sec:SIGW}}
\subsection{Formalism}
In this section, we show how the gravitational waves induced from the scalar perturbations caused by DWs. In the Newtonian gauge, the metric containing scalar perturbations $\Phi$, $\Psi$ and tensor perturbations $h_{ij}$ is
\begin{equation}
	ds^{2}=a^{2}(\tau)\Big\{-(1+2\Phi)d\tau^{2}+\big[(1-2\Psi)\delta_{ij}+\frac{1}{2}h_{ij}\big]dx^{i}dx^{j}\Big\}, 
\end{equation}
where $\tau$ is conformal time.

Assuming no anisotropic stress, we have $\Phi = \Psi$ \cite{Kodama:1984}. The tensor perturbation $h_{ij}$ is transverse-traceless (TT), and its equation of motion is given by the second-order Einstein equations Eq.~\eqref{eq:7}. Here, the prime denotes the derivative with respect to conformal time $\tau$, $\mathcal{H} = a'/a$ is the conformal Hubble parameter, and $S_{\boldsymbol{k}}$ is the source term. The polarization indices are omitted here, since both polarizations satisfy the same equation.

The source term is composed of a quadratic form of the first-order scalar perturbation $\Phi$ \cite{Baumann:2007,Ananda:2007,Acquaviva:2003,Mollerach:2004}. During the radiation-dominated era ($w = 1/3$), its expression is
\begin{equation}
	S_{\boldsymbol{k}}(\tau)=\int\frac{d^{3}q}{(2\pi)^{3/2}}e_{ij}(\boldsymbol{k})q_{i}q_{j}\Bigg[2\Phi_{\boldsymbol{q}}\Phi_{\boldsymbol{k}-\boldsymbol{q}}+\frac{4}{3(1+w)}\big(\mathcal{H}^{-1}\Phi_{\boldsymbol{q}}^{\prime}+\Phi_{\boldsymbol{q}}\big)\big(\mathcal{H}^{-1}\Phi_{\boldsymbol{k}-\boldsymbol{q}}^{\prime}+\Phi_{\boldsymbol{k}-\boldsymbol{q}}\big)\Bigg], 
\end{equation}
where $e_{ij}(\boldsymbol{k})$ is the polarization tensor ($+,\times$). This source term reflects the nonlinear coupling of scalar perturbations: two scalar modes generate a tensor source through a momentum convolution.

The equation of motion is a linear inhomogeneous equation and can be solved using the Green's function method. The Green's function $G_{\boldsymbol{k}}(\tau, \tau_1)$ satisfies
\begin{equation}
	G_{\boldsymbol{k}}^{\prime\prime}(\tau,\tau_{1})+\left(k^{2}-\frac{a^{\prime\prime}(\tau)}{a(\tau)}\right)G_{\boldsymbol{k}}(\tau,\tau_{1})=\delta(\tau-\tau_{1}). 
\end{equation}

During the radiation-dominated era the Green's function has an analytic form $G_{\boldsymbol{k}}(\tau, \tau_1) = \sin[k(\tau - \tau_1)]/k$. The solution for the tensor perturbation is
\begin{equation}
	h_{\boldsymbol{k}}(\tau)=\frac{4}{a(\tau)}\int^{\tau}d\tau_{1}\,a(\tau_{1})G_{\boldsymbol{k}}(\tau,\tau_{1})S_{\boldsymbol{k}}(\tau_{1}). 
\end{equation}

For each polarization, $\mathcal{P}_{h}(\tau,k)$ is defined by the two-point correlation function
\begin{equation}
	\langle h_{\boldsymbol{k}}^{\lambda}(\tau)h_{\boldsymbol{k}^{\prime}}^{\lambda^{\prime}}(\tau)\rangle=\delta_{\lambda\lambda^{\prime}}\delta^{(3)}(\boldsymbol{k}+\boldsymbol{k}^{\prime})\frac{2\pi^{2}}{k^{3}}\mathcal{P}_{h}(\tau,k), 
\end{equation}
substituting the solution for $h_{\boldsymbol{k}}$ and retaining the pair contractions of the scalar four-point function (with two-point spectrum $\mathcal{P}_{\Phi}$; the connected contribution is analyzed in Sec.~\ref{sec:NG}), after integrating over angles, we obtain the integral expression for $\mathcal{P}_{h}$ \cite{Kohri:2018}.

The gravitational wave energy density parameter is defined as
\begin{equation}
	\Omega_{\mathrm{GW}}(\tau,k)\equiv\frac{1}{\rho_{\mathrm{tot}}}\frac{d\rho_{\mathrm{GW}}}{d\ln k}=\frac{1}{24}\left(\frac{k}{\mathcal{H}(\tau)}\right)^{2}\overline{\mathcal{P}_{h}(\tau,k)}, 
\end{equation}
where the overline denotes the oscillation average, and the two polarization modes have been summed.

In the long-time limit $x = k\tau \to \infty$, the kernel function after oscillation averaging can be computed analytically. Before presenting the final expression, we distinguish pair-contracted and connected tensor-power contributions. The pair contractions yield Eq.~\eqref{eq:35} for the adopted two-point input, linearly related to Eq.~\eqref{eq:21}. Under the model conditions of Sec.~\ref{sec:NG}, the connected four-point contribution vanishes.

Finally, we obtain a closed-form expression for the scalar-induced gravitational wave energy spectrum during the radiation-dominated era~\cite{Xu:2020,Gao:2021}
\begin{equation}
	\begin{aligned}
		\Omega_{\mathrm{GW}}(\tau,k)=\frac{1}{12}\int_{0}^{\infty}dv\int_{|1-v|}^{1+v}du\,&\left(\frac{4v^{2}-(1+v^{2}-u^{2})^{2}}{4uv}\right)^{2}\\
		&\times\mathcal{P}_{\mathcal{R}}(ku)\mathcal{P}_{\mathcal{R}}(kv)\left(\frac{3}{4u^{3}v^{3}}\right)^{2}(u^{2}+v^{2}-3)^{2}\\
		&\times\Bigg\{\left[-4uv+(u^{2}+v^{2}-3)\ln\left|\frac{3-(u+v)^{2}}{3-(u-v)^{2}}\right|\right]^{2}\\
		&\qquad+\left[\pi(u^{2}+v^{2}-3)\Theta(u+v-\sqrt{3})\right]^2\Bigg\}, 
		\label{eq:35}
	\end{aligned}
\end{equation}
where $\mathcal{P}_{\mathcal{R}}$ is the power spectrum of the primordial curvature perturbation, and $u = |\boldsymbol{k} - \boldsymbol{p}|/k$ and $v = |\boldsymbol{p}|/k$ are dimensionless momentum variables \cite{Kohri:2018,Espinosa:2018,Domenech:2021}.

After being produced, gravitational waves redshift like radiation ($\rho_{\mathrm{GW}} \propto a^{-4}$). The formula for redshifting from the radiation-matter equality time $\tau_{\mathrm{eq}}$ to the present time $\tau_0$ is
\begin{equation}
	\Omega_{\mathrm{GW}}(\tau_{0},k)=\Omega_{\gamma,0}\left(\frac{g_{*,0}}{g_{*,\mathrm{eq}}}\right)^{\frac{1}{3}}\Omega_{\mathrm{GW}}(\tau_{\mathrm{eq}},k), 
\end{equation}
where $\Omega_{\gamma,0}$ is the photon density parameter today, and $g_{*,0}$ and $g_{*,\mathrm{eq}}$ are the numbers of relativistic degrees of freedom at the present time and at the radiation-matter equality time, respectively \cite{Baumann:2007,Watanabe:2006}. 


\subsection{Non-Gaussianity of the scalar source}\label{sec:NG}
The SIGW source used in this paper is quadratic in the scalar perturbation $\Phi$; its tensor two-point function therefore involves the scalar four-point function. We split this four-point function into its pair contractions and its connected part, following the standard treatment of non-Gaussian contributions to the induced spectrum~\cite{Adshead:2021}:
\begin{equation}
	\langle\Phi_1\Phi_2\Phi_3\Phi_4\rangle=\sum_{\rm pairings}\langle\Phi_i\Phi_j\rangle\langle\Phi_k\Phi_l\rangle+\langle\Phi_1\Phi_2\Phi_3\Phi_4\rangle_c,
	\label{eq:ng-decomp}
\end{equation}
where $\Phi_i\equiv\Phi_{\boldsymbol{k}_i}$. This decomposition is exact and does not assume Gaussianity or invoke the central-limit theorem. The pair contractions reproduce Eq.~\eqref{eq:35} when evaluated with the adopted two-point input, which is linearly related to $\mathcal{P}_{\mathcal{R}}$ in Eq.~\eqref{eq:21}; in the following we evaluate the connected part.

For the model considered here, $\Phi$ is a linear superposition of independent, spherically symmetric walls, and the connected part of its four-point function is formed from products of four wall responses that share a common nucleation event; because each wall is spherical, this function carries no azimuthal dependence.

The connected contribution to the tensor power spectrum, $\mathcal{P}^{\rm conn}_{h}(k)$, takes the form of a convolution of the connected four-point function with the source kernels~\cite{Adshead:2021}, which depend on the momenta only through their magnitudes. The two source vertices carry the polarization projections $Q_\lambda$ of the tensor $e_{ij}$ appearing in the source term above~\cite{Adshead:2021}; $\mathcal{P}^{\rm conn}_{h,\lambda}(k)$ denotes the corresponding contribution of polarization $\lambda$. With $\boldsymbol{k}$ chosen as the polar axis, the projection is defined by
\begin{equation}
	Q_\lambda(\boldsymbol{k},\boldsymbol{q})\equiv e_\lambda^{ij}(\hat{\boldsymbol{k}})\,q_iq_j,
\end{equation}
where $e_\lambda^{ij}$ are the $+,\times$ polarization components of $e_{ij}$. In the standard $+,\times$ basis, $Q_+\propto\sin^2\theta\cos 2\phi$ and $Q_\times\propto\sin^2\theta\sin 2\phi$, so that
\begin{equation}
	\sum_{\lambda=+,\times}Q_\lambda(\boldsymbol{k},\boldsymbol{q}_1)Q_\lambda(\boldsymbol{k},\boldsymbol{q}_2)\propto \sin^2\theta_1\sin^2\theta_2\cos\!\bigl[2(\phi_1-\phi_2)\bigr].
	\label{eq:ng-polarization-sum}
\end{equation}
All other factors in the connected contribution are independent of $\phi_1$ and $\phi_2$. The azimuthal integration therefore gives
\begin{equation}
	\int_0^{2\pi}d\phi_1\,\cos\!\bigl[2(\phi_1-\phi_2)\bigr]=0\quad\Longrightarrow\quad \mathcal{P}^{\rm conn}_{h,\lambda}(k)=0\ \ (\lambda=+,\times),\qquad \mathcal{P}^{\rm conn}_{h}(k)=0.
	\label{eq:ng-vanishing}
\end{equation}
The same cancellation holds separately in the $+$ and $\times$ channels, rather than arising only after their sum. Thus, for this model, the connected part of the four-point function gives no additional contribution to the induced spectrum.


\subsection{The model and numerical results}
In this section, we construct an extended-period nucleation model in which the DW tension $\sigma(t)$ varies smoothly over an extended period, allowing DWs to acquire a distribution of comoving radii.

We investigate a two-field inflationary model with the inflaton $\phi$ and a spectator field $\chi$. The action is
\begin{equation}
	S=\int d^4 x\sqrt{-g}\left[-\frac{M_{\text{p}}^2}{2} R+\frac{1}{2}\partial_\mu\phi\partial^\mu\phi+\frac{1}{2}\partial_\mu\chi\partial^\mu\chi+V(\phi,\chi)\right],
\end{equation}
with the potential $V(\phi, \chi)$,
\begin{equation}
	V(\phi,\,\chi) = f(\phi) + \frac{\lambda_{\chi}}{4} \left[ \chi^2 - \Lambda^2 \left( 1 - \mathrm{e}^{ -\frac{(\phi - \phi_c)^2}{\Delta^2} } \right) - m^2 \right]^2. 
	\label{eq:4}
\end{equation}

This potential features two degenerate vacua in the $\chi$-direction. Since $\chi$ is a spectator field during inflation—the field responsible for domain wall formation via quantum tunneling—the inflationary dynamics are governed solely by $f(\phi)$, which is chosen to be consistent with Planck 2018 constraints~\cite{Planck:2018}:
\begin{equation}
	f(\phi) = a \left( 1 - \mathrm{e}^{-A \phi} \right)^2 , 
\end{equation}
with $a=1.09\times10^{-10}$ and $A=0.91$, yielding $n_s=0.9610$, $r=0.0035$, and $\ln(10^{10}A_S)=3.0319$.

The Friedmann equation and the equations of motion for $\phi$ and $\chi$ are
\begin{equation}
	\begin{aligned}
		&H^2 = \frac{1}{3M_{\text{p}}^2}\left(\frac{1}{2}\dot{\phi}^2+\frac{1}{2}\dot{\chi}^2+V(\phi,\chi)\right), \\
		&\ddot{\phi} + 3H\dot{\phi} + \frac{\partial V}{\partial\phi} = 0, \\
		&\ddot{\chi} + 3H\dot{\chi} + \frac{\partial V}{\partial\chi} = 0.  
	\end{aligned}
	\label{eq:system}
\end{equation}

The three parameter sets used in our computation are listed in Table~\ref{tab:1}.

\begin{table}[htbp]
	\centering
	\setlength{\tabcolsep}{8pt}
	\begin{tabular}{c c c c c c}
		\toprule
		Set & $\phi_c/M_{\text{p}}$ & $\lambda_\chi$ & $\Delta/M_{\text{p}}$ & $m/M_{\text{p}}$ & $\Lambda/M_{\text{p}}$ \\
		\hline
		1   & 3.18  & 185.64  & 2.0  & $2.5155 \times 10^{-6}$ & $7.71733 \times 10^{-6}$ \\
		2   & 3.33087  & 113.11  & 1.17  & $3.2456 \times 10^{-6}$ & $6.25010 \times 10^{-6}$ \\
		3   & 3.18  & 61.422  & 4.0  & $4.3810 \times 10^{-6}$ & $1.30776 \times 10^{-5}$ \\
		\toprule
	\end{tabular}
	\caption{The parameter sets.}
	\label{tab:1}
\end{table}

The DW tension derived from Eq.~\eqref{eq:4} is
\begin{equation}
	\sigma(t) = \frac{4}{3} \sqrt{\frac{\lambda_\chi}{2}} \left[ \Lambda^2 \left( 1 - \mathrm{e}^{ -\frac{ \left( \phi(t) - \phi_c \right)^2 }{\Delta^2} } \right) + m^2 \right]^{\frac{3}{2}}, 
	\label{eq:sigma}
\end{equation}

\begin{figure}[htbp]
	\centering
	\includegraphics[width=0.8\textwidth]{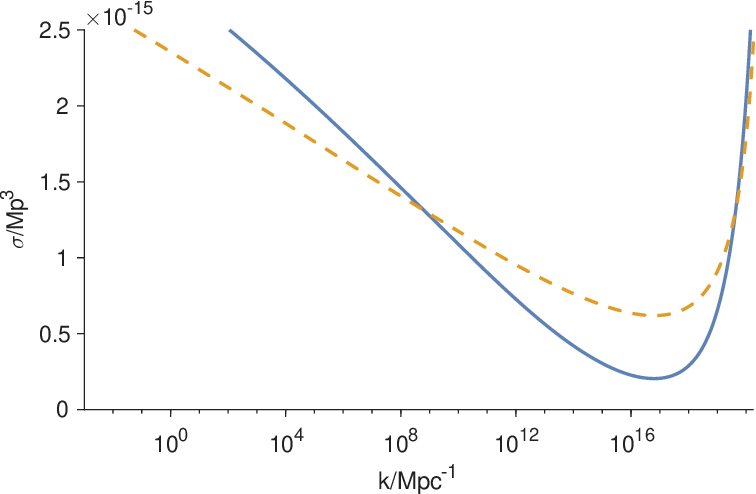}
	\caption{The comparison of the $\sigma$ profile width between Parameter Set 1 (blue solid line) and Parameter Set 3 (orange dashed line).}
	\label{fig:sigma}
\end{figure}

Unlike the spike-like profile in the small-period nucleation model, $\sigma(t)$ in Eq.~\eqref{eq:sigma} varies smoothly because the Gaussian factor $e^{-(\phi-\phi_c)^2/\Delta^2}$ provides a controlled width governed by the parameter $\Delta$. A larger $\Delta$ broadens the $\sigma$ profile, extending the nucleation period and producing DWs with a distribution of comoving radii ($\gamma>1$), as shown in Fig.~\ref{fig:sigma}. 

\begin{figure}[htbp]
	\centering
	\includegraphics[width=0.8\textwidth]{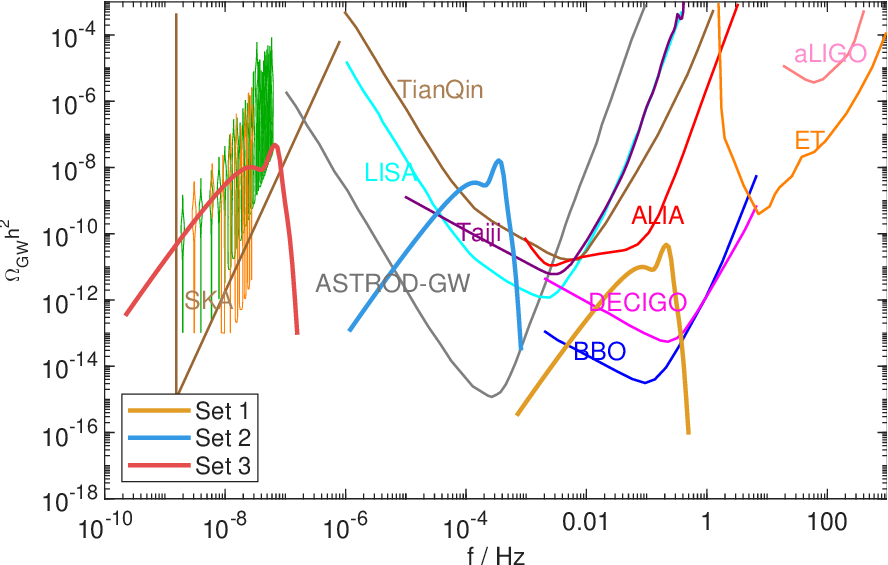}
	\caption{Predicted energy spectra of scalar-induced GWs with the parameter Set 1 (yellow), Set 2 (blue), and Set 3 (red) in Table~I. The sensitivity curves of the GW detectors are shown, including SKA~\cite{SKA:2020}, ASTROD-GW~\cite{ASTROD-GW:2013},  LISA~\cite{LISA:2017,Robson:2019,AmaroSeoane:2023}, Taiji~\cite{Taiji:2020}, TianQin~\cite{TianQin:2016}, DECIGO~\cite{Kawamura:2011,Seto:2001,Yagi:2011}, BBO~\cite{BBO:2006}, ALIA~\cite{ALIA:2011}, aLIGO~\cite{aLIGO:2015}, and ET~\cite{ET:2010},  The orange region shows the data of EPTA~\cite{EPTA:2023,CPTA:2023}, the green
		region shows the PTA data of NANOGrav~\cite{NANOGrav:2023,NANOGrav:2023a}, which are summarized in Ref.~\cite{Moore:2015,Kuroda:2015,Schmitz:2021}. }
	\label{fig:2}
\end{figure}

Solving the background dynamics and computing $\gamma$, $k_{\text{cut}}$, and $\bar{\rho}_{\text{DW}}/\rho_r$ for each parameter set, we evaluate the curvature power spectrum Eq.~\eqref{eq:21} and the SIGW spectrum. The present-day gravitational-wave energy density $\Omega_{\mathrm{GW}}(f)$ for the three parameter sets is shown in Fig.~\ref{fig:2}.

\begin{table}[htbp]
	\centering
	\setlength{\tabcolsep}{8pt}
	\begin{tabular}{l c c c}
		\toprule
		  & Set 1 & Set 2 & Set 3 \\
		\hline
		$H(t_e)$              & $4.393\times10^{-6}$  & $4.393\times10^{-6}$  & $4.393\times10^{-6}$ \\
		$\sigma(t_e)$         & $3.284\times10^{-15}$ & $3.275\times10^{-15}$ & $2.749\times10^{-15}$ \\
		$\Delta t$            & $1.861$               & $1.554$               & $2.286$              \\
		$\gamma$              & $1.916\times10^{16}$  & $4.559\times10^{27}$  & $9.575\times10^{53}$  \\
		$n$                   & $6.687\times10^{41}$  & $3.006\times10^{33}$  & $2.133\times10^{22}$   \\
		$k_{\rm cut}$         & $1.410\times10^{14}$  & $2.326\times10^{11}$  & $4.471\times10^{7}$    \\
		$\bar{\rho}_{\rm DW}/\rho_r$ & $1.974\times10^{-10}$ & $1.752\times10^{-15}$ & $1.584\times10^{-28}$ \\
		$p$                   & $2.914\times10^{-9}$  & $2.106\times10^{-15}$ & $2.658\times10^{-28}$ \\
		$S_E(min)$            & 21.77                 & 35.73                 & 65.82              \\
		$\delta_wH(max)$  & 0.2352879             & 0.2352828             & 0.2352584             \\
		\toprule
	\end{tabular}
	\caption{The key quantities for the three benchmark sets in Table~\ref{tab:1}. All dimensionful quantities are in units of the Planck mass $M_{\rm p}$; $n$ and $k_{\rm cut}$ are comoving quantities in units of ${\rm Mpc}^{-3}$ and ${\rm Mpc}^{-1}$; $\Delta t$ is the duration of the nucleation period in e-folds; $\bar\rho_{\rm DW}/\rho_r$ is evaluated after the conversion of the DWs; $p$ is the nucleation probability per Hubble volume Eq.~\eqref{eq:p}; $S_E(min)$ is the minimum of the Euclidean action Eq.~\eqref{eq:SE} over the nucleation period; $\delta_wH(max)$ is the maximum dimensionless wall thickness over the nucleation period.}
	\label{tab:2}
\end{table}

We verify that $p < 1$ for all three sets, satisfying the consistency condition discussed in Sec.~\ref{sec:model}. The spectral shape for all three sets follows the characteristic form of scalar-induced GWs from a steep power-law curvature power spectrum: a low-frequency tail $\Omega_{\mathrm{GW}} \propto f^{3}$ (dictated by the $k^{3}$ slope of $\mathcal{P}_{\mathcal{R}}$)  and a rapid cutoff above the peak due to the truncation at $k_{\mathrm{cut}}$.The predicted energy spectrum for Set~1 (yellow curve) peaks at $f_{\mathrm{peak}}\sim 2.1\times10^{-1}\,\mathrm{Hz}$ with an amplitude $\Omega_{\mathrm{GW}}h^{2}\sim 4.6\times10^{-11}$, falling within the sensitivity band of DECIGO and BBO. Set~2 (blue curve) has a peak at $f_{\mathrm{peak}}\sim 3.5\times10^{-4}\,\mathrm{Hz}$ and $\Omega_{\mathrm{GW}}h^{2}\sim 1.6\times10^{-8}$, making it accessible to LISA/Taiji in the millihertz range. Set~3 (red curve), the central result of this work, exhibits a broad peak centered at $f_{\mathrm{peak}}\sim 6.7\times10^{-8}\,\mathrm{Hz}$ with a peak amplitude $\Omega_{\mathrm{GW}}h^{2}\sim 4.8\times10^{-8}$. This signal lies directly inside the confidence region of the common-spectrum process reported by NANOGrav and EPTA thereby providing a viable explanation for the detected nanohertz stochastic gravitational-wave background. The tunability of the GW spectrum across frequency bands can be understood from the interplay between $\gamma$ and $k_{\text{cut}}$. The width parameter $\Delta$ controls how broadly DWs nucleate during inflation, which determines the distribution of their comoving radii. A broader distribution yields $\gamma > 1$, directly enhancing the amplitude of the curvature power spectrum $\mathcal{P}_\mathcal{R}(k) \propto \gamma(\bar{\rho}_{\text{DW}}/\rho_r)^2(k/k_{\text{cut}})^3$. Meanwhile, $k_{\text{cut}}$ is set by the mean DW separation and can be tuned by adjusting the potential parameters. With $\gamma$ controlling the amplitude and $k_{\text{cut}}$ setting the peak frequency, the model can produce viable GW signals from the nanohertz band (PTA) up to the decihertz band (DECIGO/BBO).

Moreover, when the scalar perturbations near $k_{\text{cut}}$ re-enter the horizon, they will produce the primordial black holes (PBHs) through gravitational collapse~\cite{Deng:2017,Tanahashi:2015}. Thus we also calculate the abundance of PBHs using the Press-Schechter approach of gravitational collapse~\cite{Press:1974,Carr:1975}, and find that for Set~3, the peak mass of PBHs is around $4.267 \times 10^{-3}\,M_\odot$ and the fraction in dark matter is about $4.425 \times 10^{-5}$, which are not overproduced. For Set~1 and Set~2, the mass abundance of PBHs is even smaller and can be neglected.

\section{Summary \label{sec:summary}}
We have presented a kind of DW nucleation model with an extended-period nucleation during inflation and demonstrated that it can account for the nanohertz stochastic gravitational-wave background observed by PTA collaborations. One example is based on a two-field inflationary potential (Eq.~\eqref{eq:4}) in which the DW tension $\sigma(t)$ evolves smoothly over an extended period, controlled by the width parameter $\Delta$. This leads to a distribution of DW radii, captured by the factor $\gamma>1$, which enhances the curvature power spectrum. Using Poisson statistics, we derived the curvature perturbation power spectrum $\mathcal{P}_{\mathcal{R}}(k)\propto k^3$ with a cutoff at $k_{\mathrm{cut}}$, as given in Eq.~\eqref{eq:21}.

Applying the scalar-induced gravitational-wave formalism, we computed the present-day gravitational-wave energy density for three parameter sets (TABLE~\ref{tab:1}). Among them, the result for Parameter Set~3 is relatively important. This signal lies within the confidence region of the NANOGrav common-spectrum process and is also compatible with the upper limits from EPTA, thereby providing a natural explanation for the detected nanohertz background. The extended-period nucleation mechanism offers two crucial advantages: (i) the enhancement factor $\gamma>1$ boosts the curvature power spectrum without requiring an overly large DW energy fraction; (ii) the cutoff scale $k_{\mathrm{cut}}$ can be tuned via the potential width $\Delta$, enabling the same model to produce signals in the nanohertz (Set~3), millihertz (Set~2), and hertz (Set~1) bands. By selecting different parameters, the model can produce signals detectable by different gravitational-wave detectors.

In summary, our work highlights the importance of the nucleation-time window in determining the observational signatures of domain-wall-induced gravitational waves. The extended-period nucleation DW model provides a more realistic description of the nucleation process and observationally viable explanation for the nanohertz stochastic gravitational-wave background discovered by PTAs. Future PTA experiments will further test the spectral shape, while space-based interferometers such as LISA, Taiji, and DECIGO, also BBO can probe the higher-frequency branches of the model. The possible connection to primordial black holes offers an additional avenue for multi-messenger verification. 

Finally, we note that, in addition to quantum tunneling, DWs can also form during inflation through stochastic diffusion of the spectator field \cite{LindeLinde:1994,Gonzalez:2023,Faria:2025}. Walls produced by this mechanism are not generically spherical, and their non-spherical collapse could potentially yield a larger gravitational-wave signal than the spherical configurations considered here. A quantitative comparison between the two nucleation channels, including the relative contribution of the stochastic mechanism to the nanohertz SGWB, is left for future work.

\begin{acknowledgments}
This work was supported by “the Natural Science Basic Research Program of Shaanxi
Province” No. 2023-JC-YB-072. And supported by “the Fundamental Research Funds for
the Central Universities” No. ZYTS25130.
\end{acknowledgments}

\end{document}